\documentclass[epj, final]{svjour}
\usepackage{graphics}
\usepackage{amsmath}
\begin{document}
\title{Hydrogenated pyrene: Statistical single-carbon loss below the knockout threshold}
\author{M. Wolf\inst{1}\and L. Giacomozzi\inst{1}\and M. Gatchell\inst{1}\and N. de Ruette\inst{1}\and M. H. Stockett\inst{1,2}\and H. T. Schmidt\inst{1}\and H. Cederquist\inst{1}\and H. Zettergren\inst{1}
}                     
%
%
\institute{Department of Physics, Stockholm University, Stockholm, SE--106 91, Sweden,  \and Department of Physics and Astronomy, Aarhus University, DK--8000 Aarhus, Denmark}
\mail{michael.wolf@fysik.su.se}
\date{Received: date / Revised version: date}
%
\abstract{An ongoing discussion revolves around the question of what effect hydrogenation has on carbon backbone fragmentation in Polycyclic Aromatic Hydrocarbons (PAHs). In order to shed more light on this issue, we have measured absolute single carbon loss cross sections in collisions between native or hydrogenated pyrene cations (C$_{16}$H$_{10+m}^{+}, m$ = 0, 6, 16) and He as functions of center-of-mass energies all the way down to 20~eV. Classical Molecular Dynamics (MD) simulations give further insight into energy transfer processes and also yield $m$-dependent threshold energies for prompt (femtoseconds) carbon knockout. Such fast, non-statistical fragmentation processes dominate CH$_x$-loss for native pyrene ($m$=0), while much slower statistical fragmentation processes contribute significantly to single-carbon loss for the hydrogenated molecules ($m=6$ and $m=16$). The latter is shown by measurements of large CH$_x$-loss cross sections far below the MD knockout thresholds for C$_{16}$H$_{16}^{+}$ and C$_{16}$H$_{26}^{+}$.
} 
\maketitle
\section{Introduction}
\label{intro}

Polycyclic Aromatic Hydrocarbons (PAHs) are believed to be present throughout the interstellar medium and to contribute to the so called Aromatic Infrared Bands (AIBs) in a range of astrophysical objects \cite{Allamandola1999,Tielens2008,Tielens2013}. Production and destruction mechanisms of PAHs in astrophysical environments are therefore of large interest and have been the subject of recent intense theoretical and experimental research \cite{Tielens2013,Denifl2005,Denifl2006,Rauls2008,Micelotta2010a,Micelotta2010b,Postma2010,Holm2011,Johansson2011,Martin2012,Martin2013,Reitsma2013,Postma2014,C4CP03293D,Mishra2014,West2014,Kaiser2015,Delaunay2015}.
In space, PAHs may be excited through photo-absorption processes, or in collisions with electrons or heavy particles (ions/atoms). In many such cases there is enough time (picoseconds) for the internal energy to distribute over all vibrational degrees of freedom before the PAHs cool down through emission of photons, electrons, and/or fragments. 
Then any fragmentation process will be \textit{statistical}, favouring the lowest-energy dissociation channels which are H- and/or C$_2$H$_2$-loss for native PAHs \cite{Martin2012,Gatchell2015}. 

In addition, close interactions such as those between a colliding He atom and individual atoms in the PAH may transfer enough energy for prompt single-atom knockout. This occurs on femtosecond timescales, long before the energy has had time to distribute statistically over the molecular system and may therefore be referred to as \textit{non-statistical} fragmentation \cite{Chen2014,Gatchell2014b,Stockett2014,Stockett2015,Stockett2015b}. Here one should note that the present definitions of statistical and non-statistical  processes relate to the fragmentation step only. The distributions of impact parameters in the exciting atom-molecule collisions are random in both cases. The energy transfers in these collisions must be substantially greater than the relevant fragmentation barriers for the decay to proceed within microseconds (the present experimental timescale). The so produced fragments will then themselves carry substantial excitation energies, which may lead to secondary statistical fragmentation processes  \cite{Stockett2014,Stockett2015}.

\begin{figure}
\centering
\resizebox{3in}{!}{%
  \includegraphics{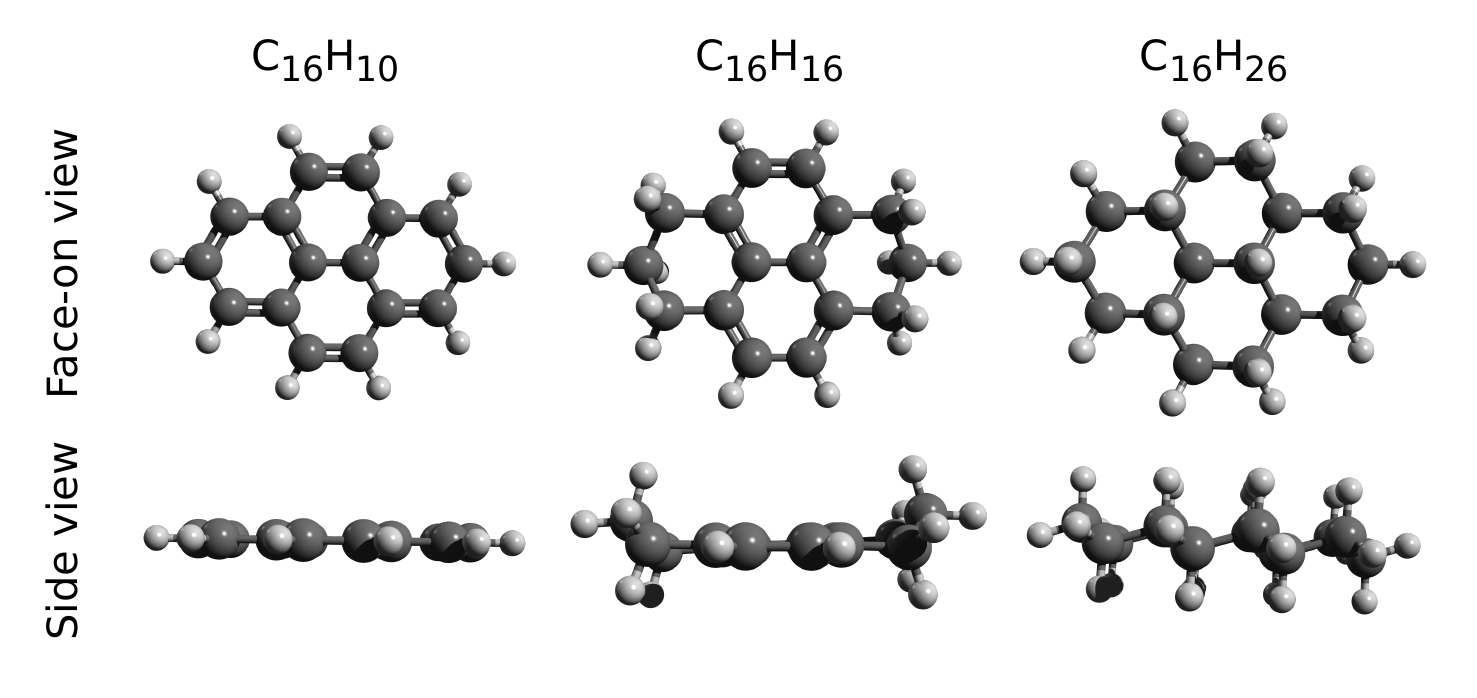}
}
\caption{Molecular structures of pyrene (C$_{16}$H$_{10}$, left), hexahydropyrene (C$_{16}$H$_{16}$, center) and hexadecahydropyrene (C$_{16}$H$_{26}$, right). The native pyrene is completely planar with only aromatic sp$^2$-type bonds while hydrogenation leads to weaker sp$^3$-type single bonds and non-planar structures. Top row: Face-on views of the three molecules. Bottom row: Side views of the molecules.
}
\label{fig: pyrene} 
\end{figure}

A very interesting issue concerns the study of the effect of hydrogenation on the stability of the PAH carbon backbone. Hydrogenation of PAHs results in two counteracting effects. On one hand, the related conversion of unsaturated bonds to single bonds weakens the carbon backbone of the molecule (see Fig. \ref{fig: pyrene}). On the other hand, the binding energies for the additional hydrogen atoms are smaller than for the native ones and depend on the degree of hydrogenation \cite{Cazaux2016}. Loosing these additional H-atoms may then cool the remaining PAH system.

Reitsma \textit{et al.} recently presented an experimental study on H-loss from native and hydrogenated coronene (C$_{24}$H$_{12+m}$, $m$ = 0--7) following soft X-ray absorption \cite{Reitsma2014}. They found that the hydrogenated molecule lost fewer of its native hydrogen atoms than native coronene ($m=0$) for a given photon energy. They also found that this effect increased with the degree of hydrogenation and concluded that hydrogenation has a net protective effect on the carbon backbone for PAHs of any size \cite{Reitsma2014}. However, Gatchell \textit{et al.} \cite{Gatchell2015} later found that hydrogenation does not always protect PAHs from fragmentation. They showed that hydrogenation leads to larger probabilities for carbon backbone fragmentation in collisions between C$_{16}$H$_{10+m}^+$ ($m$ = 0, 6, and 16) and He \cite{Gatchell2015}. 

Here, we present experimental results for collisions between native ($m=0$) or hydrogenated ($m>0$) pyrene cations (see Fig. \ref{fig: pyrene}) and He for a wide range of center-of-mass energies of 20--200~eV in order to investigate the threshold behaviours of the fragmentation processes in detail. From the mass spectra and primary beam attenuation measurements we extract absolute backbone and CH$_x$-loss cross sections as functions of energy. In addition, we have performed classical MD simulations of collisions between C$_{16}$H$_{10+m}$ ($m$ = 0, 6, 16) and He in the time range of 100 femtoseconds. The experimental and MD-results show that native and hydrogenated pyrene molecules fragment very differently in collisions with He atoms for collision energies below 50~eV. This is fully consistent with the idea that the weakening of the carbon backbone is a much more important effect than the increase in heat capacity as pyrene is hydrogenated.

\section{Experimental Details}

\begin{figure}
\centering
\resizebox{3in}{!}{%
  \includegraphics{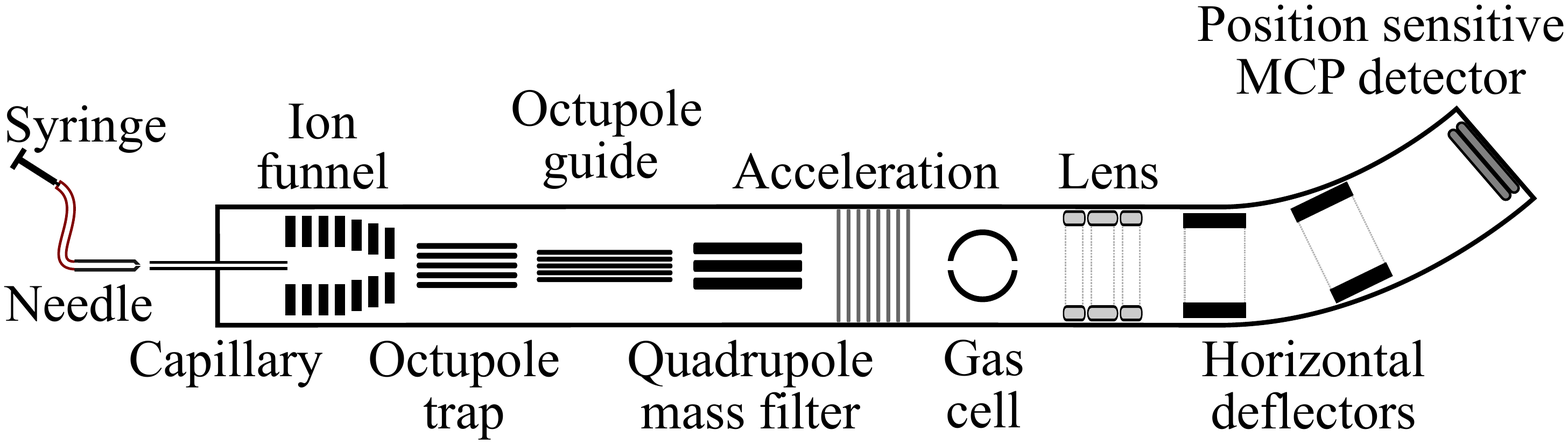}
}
\caption{Schematic of the experimental setup.}
\label{fig: Lab_schematic} 
\end{figure}
The measurements were conducted with the setup shown in Fig. \ref{fig: Lab_schematic}, which is a part of the DESIREE facility \cite{Schmidt2008,Thomas2011} at Stockholm University. All three molecules (Fig \ref{fig: pyrene}) were purchased from Sigma-Aldrich. The molecules are dissolved separately together with silver nitrate in methanol and dichloromethane and ionized in an ElectroSpray Ionization (ESI) source. The ions from the ESI source are transported through a heated capillary into an ion funnel. After the ion funnel the ions are guided through a beamline consisting of an octupole ion trap, an octupole ion guide and a quadrupole mass filter. Then the mass-selected ions enter the acceleration stage between the potential of the high voltage platform, which can be varied between 0.8 and 10 kV, and the interaction region with a gas cell on ground potential. For the molecules studied here this results in center-of-mass energies of 20--200~eV for collisions with He. The ions collide with the helium target in the 4~cm long gas cell at adjustable pressure. The pressure is measured by means of a capacitance manometer. A set of deflectors is used to separate the different charged fragments before they hit a position sensitive, 40~mm diameter, double-stack MCP detector. A more extensive description of the experimental technique and setup can be found in Ref. \cite{Haag2011}.

\section{Simulation Details}
\label{sec:MD}
We have performed simulations of collisions between He and C$_{16}$H$_{10+m}$, $m$ = 0, 6, 16, using classical Molecular Dynamics (MD) as implemented in the LAMMPS software package \cite{Plimpton:1995aa,lammps_web}. In these simulations we have used the reactive Tersoff potential \cite{PhysRevB.37.6991,PhysRevB.39.5566} to model the intramolecular bonds, and the breaking of C-C and C-H bonds. The atomic parameters used in the Tersoff potential for C atoms were obtained from Ref. \cite{PhysRevB.39.5566} and the parameters for H are from Ref. \cite{:/content/aip/journal/jap/86/4/10.1063/1.370977}. The mixing terms for C-H interactions are from Ref. \cite{C4CP03293D}. Interactions between the He projectile and the individual C and H atoms in the molecules are modelled using the Ziegler-Biersack-Littmark (ZBL) potential \cite{zbl_pot_book}. The ZBL potential models Rutherford-like elastic scattering of screened atomic nuclei, which is the dominant mechanism for energy transfer in the present simulations for collisions of 15--200~eV \cite{Chen2014}.

\begin{figure}
\centering
\resizebox{3in}{!}{%
  \includegraphics{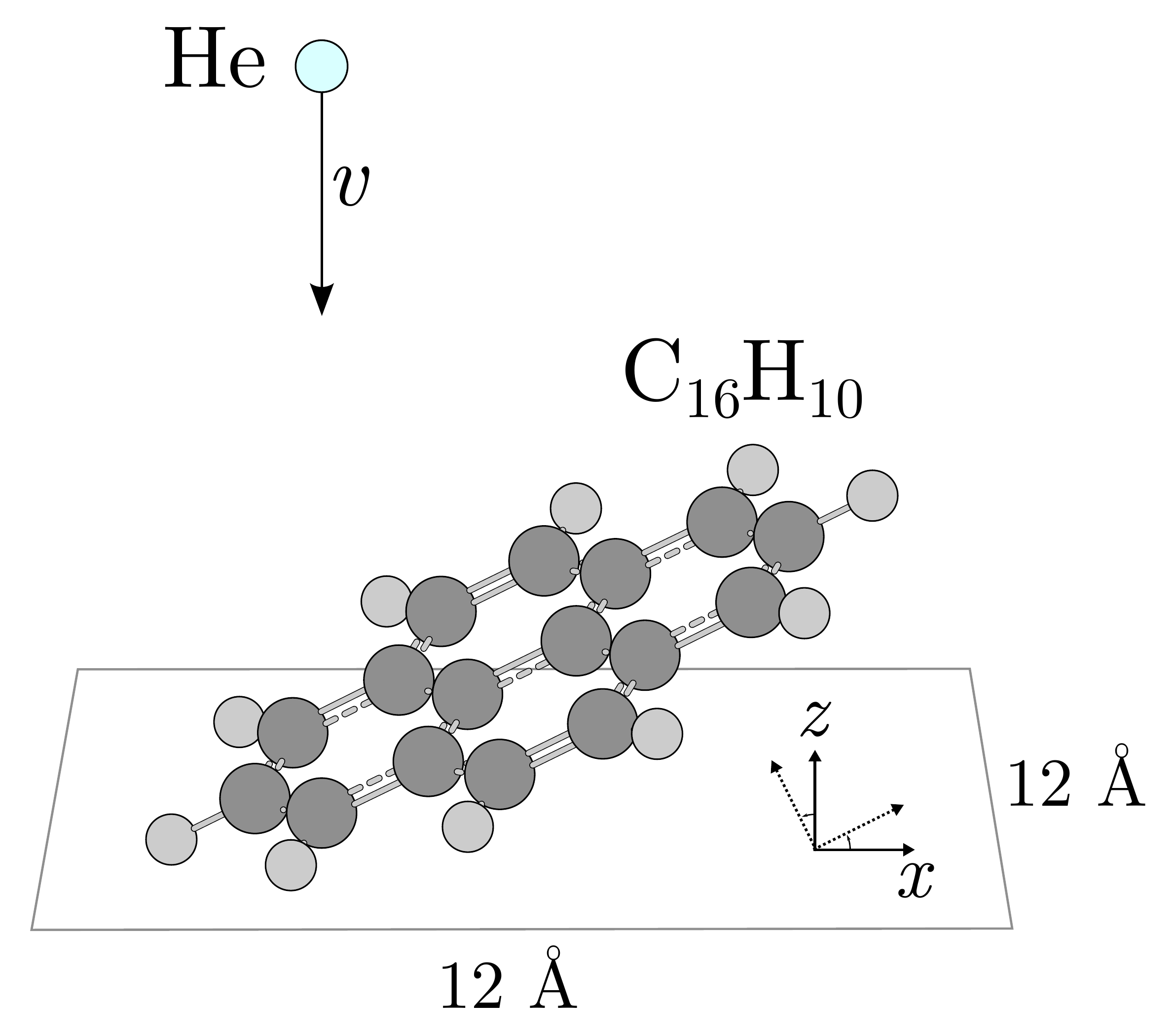}
}
\caption{Schematic of the MD simulations (see text).}
\label{fig: simulation} 
\end{figure}

Individual collision events are simulated in the following way: The optimized (using the Tersoff potential) molecule (C$_{16}$H$_{10+m}$, $m$ = 0, 6, or 16)  is positioned with its center-of-mass at the origin of a coordinate system. For each collision the molecule is randomly rotated in three dimensions around the origin. The He projectile is then launched from a randomly generated position in the $z = 10$~\AA\ plane within a square that extends out to $x = \pm 6$~\AA\ and $y = \pm 6$~\AA\ as indicated in Fig. \ref{fig: simulation}. The He atom follows a trajectory in the negative $z$-direction with a velocity set to obtain the desired energy in the He+PAH center-of-mass system. Using a time step of $10^{-17}$~s, each individual trajectory is followed for 100 femtoseconds in 10,000 steps. At the end of the simulation the positions and velocities of all atoms (the He atom and all the C and H atoms) are recorded and analysed for bond cleavage. This procedure is repeated for 10,000 random trajectories for each collision energy and for each one of the three molecules in Fig. \ref{fig: pyrene}.

With these MD simulations we are able to study fragmentation on femtosecond timescales and distributions of energy transfers to the molecular systems. There are two aspects to take note of when comparing the experimental results with the simulations. First, we only consider interactions with the atomic nuclei in the simulations---not scattering on the electrons. This means that the real energy transfers are larger than those calculated here at least for the higher energies in our range. Second, the simulations are for neutral molecules while the experiments are for cations. The difference in charge should, however, have little influence on the prompt fragmentation processes as shown earlier \cite{Chen2014}.  

\section{Results and Discussion}
\label{sec: ResDis}

\subsection{Fragmentation mass spectra of C$_{16}$H$_{10}^+$, C$_{16}$H$_{16}^+$, and C$_{16}$H$_{26}^+$}
\begin{figure*}
\centering
\resizebox{0.85\textwidth}{!}{%
\includegraphics{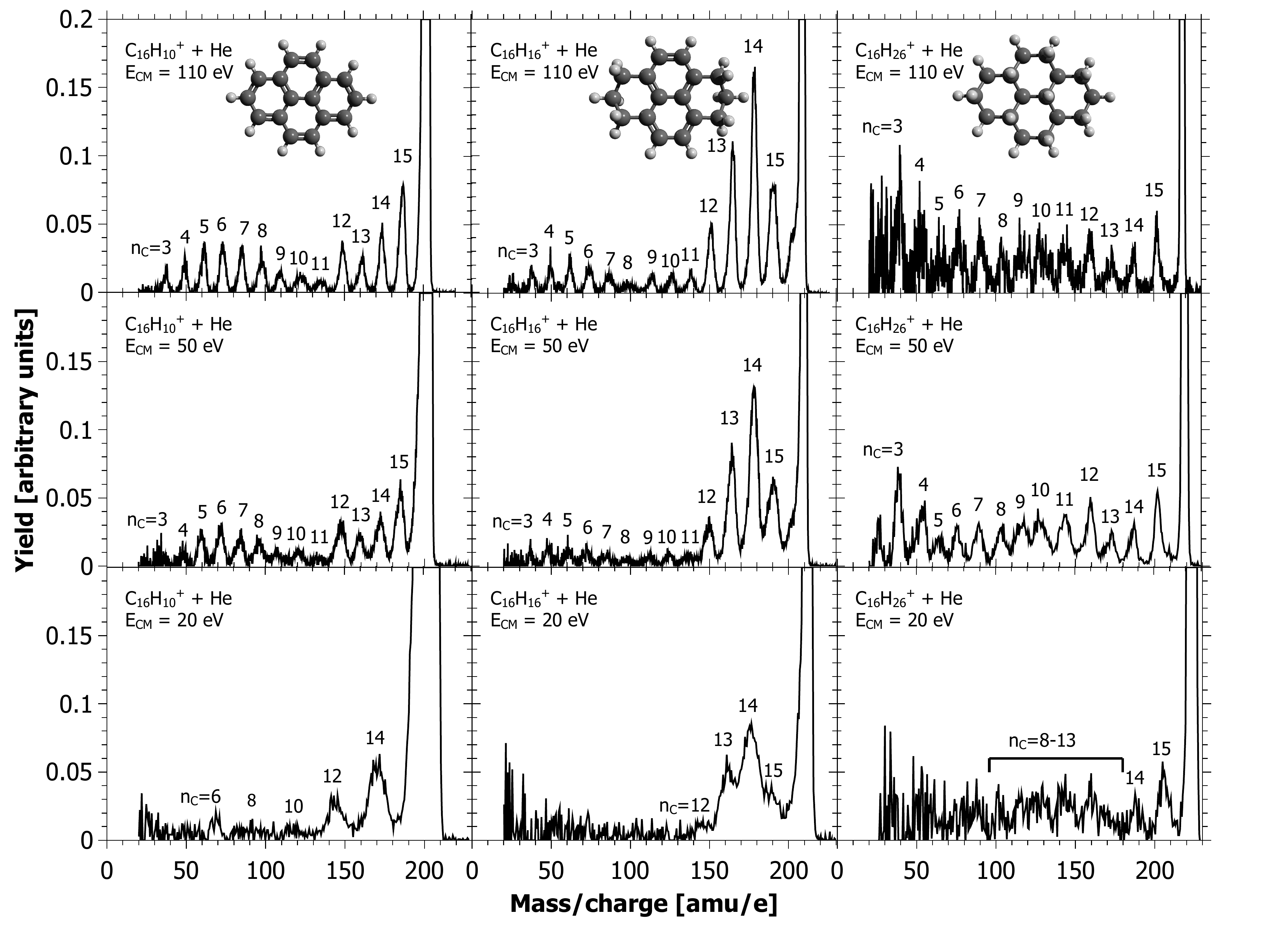}
}
\caption{Mass spectra for collisions between C$_{16}$H$_{10+m}^+$ ($m=0, 6, 16$) and helium at $E_{CM} = 110$~eV (top row), $E_{CM} = 50$~eV (center row) and $E_{CM} = 20$~eV (bottom row). The labels give the number of carbon atoms, n$_\text{C}$, in the fragment peaks. The insets in the top row shows the molecular structures for $m = 0$ (native pyrene), $m = 6$ (6 additional H-atoms) , and $m = 16$ (16 additional H atoms).}
\label{fig: MassSpectra}       
\end{figure*}

In Fig. \ref{fig: MassSpectra} we show experimental mass spectra from collisions between He and C$_{16}$H$_{10+m}^+$ with, from left to right, $m$ = 0, 6, and 16, at center-of-mass energies of $E_{CM}$ = 110, 50, and 20~eV.  These spectra show a dominant peak at the mass of the intact molecule (at 202, 208 and 218 amu, respectively). 

For 110~eV C$_{16}$H$_{10}^+$+He collisions, single carbon knockout gives the CH$_x$-loss peak with 15 remaining carbon atoms in a singly charged fragment \cite{C4CP03293D,Gatchell2015}. This is a unique fingerprint of a non-statistical fragmentation process. Statistical fragmentation of the heated molecules mainly leads to H- and/or C$_2$H$_2$-loss, which have dissociation energies of about 5~eV \cite{Martin2012,Gatchell2015}. Secondary fragmentation processes may follow after a primary statistical C$_2$H$_2$-loss or after primary non-statistical CH$_x$-loss. Such secondary fragmentation steps are mainly statistical and may result in additional losses of, e.g., C$_2$H$2$-units resulting in fragments with 12 or 13 carbon atoms, respectively. For high internal excitation energies, smaller fragments with less than 12 carbon atoms may be produced.    
At 110~eV the yield of fragments containing 13 and 14 C atoms increases while the yield of fragments with less than 11 C atoms decreases when going from C$_{16}$H$_{10}^+$ to C$_{16}$H$_{16}^+$ (left and middle top panels in Fig. \ref{fig: MassSpectra}). This would indicate that there is some cooling effect due to hydrogenation. Yet as it has been shown before, the \textit{total} carbon backbone fragmentation cross section increases between C$_{16}$H$_{10}^+$ and C$_{16}$H$_{16}^+$ \cite{Gatchell2015}. The mass spectrum of fully hydrogenated C$_{16}$H$_{26}^+$ at 110~eV displays a very different---much flatter---fragment distribution, which most likely is due to the conversion of all aromatic C-C bonds to single bonds  \cite{Gatchell2015}. 


As can be seen in the bottom left panel of Fig. \ref{fig: MassSpectra}, 20~eV C$_{16}$H$_{10}^+$+He collisions do not lead to direct single carbon knockout and fragment peaks with odd numbers of carbon atoms are no longer visible. The completely dominant fragmentation mechanism is then statistical fragmentation of internally heated molecules, which---as discussed above---predominantly occurs through H- and/or C$_2$H$_2$-losses \cite{Martin2012,Gatchell2015}. The decrease in transferred energy between 50~eV and 20~eV (middle and lower rows in Fig. \ref{fig: MassSpectra}) suppresses multiple secondary fragmentation steps for all three molecules and leads to increases in the yields of fragments with 14 and 12 carbon atoms while lighter fragments become less important especially for C$_{16}$H$_{10}^+$ and C$_{16}$H$_{16}^+$. Note that there still is a clear signal of fragments with odd numbers of carbon atoms (13 and 15) for C$_{16}$H$_{16}^+$ at 20~eV. 

The spectrum for 20~eV C$_{16}$H$_{26}^+$+He collisions in the lower right panel exhibits a flat intensity distribution without any clearly dominating peaks. Here we also still see fragments with even and odd numbers of carbon atoms. This can be understood as the weakening of the carbon backbone through hydrogenation gives dissociation energies as low as 1.60~eV for CH$_{3}$-loss and 2.02~eV for H-loss for C$_{16}$H$_{26}^+$ in recent density functional theory calculations \cite{Gatchell2015}. These energies are lower than those for C$_2$H$_x$-loss \cite{Gatchell2015} and the initial steps of carbon backbone fragmentation will then often be \textit{statistical} CH$_x$-loss.

\subsection{Absolute CH$_x$-loss cross sections}

\begin{figure}
\resizebox{3in}{!}{%
  \includegraphics{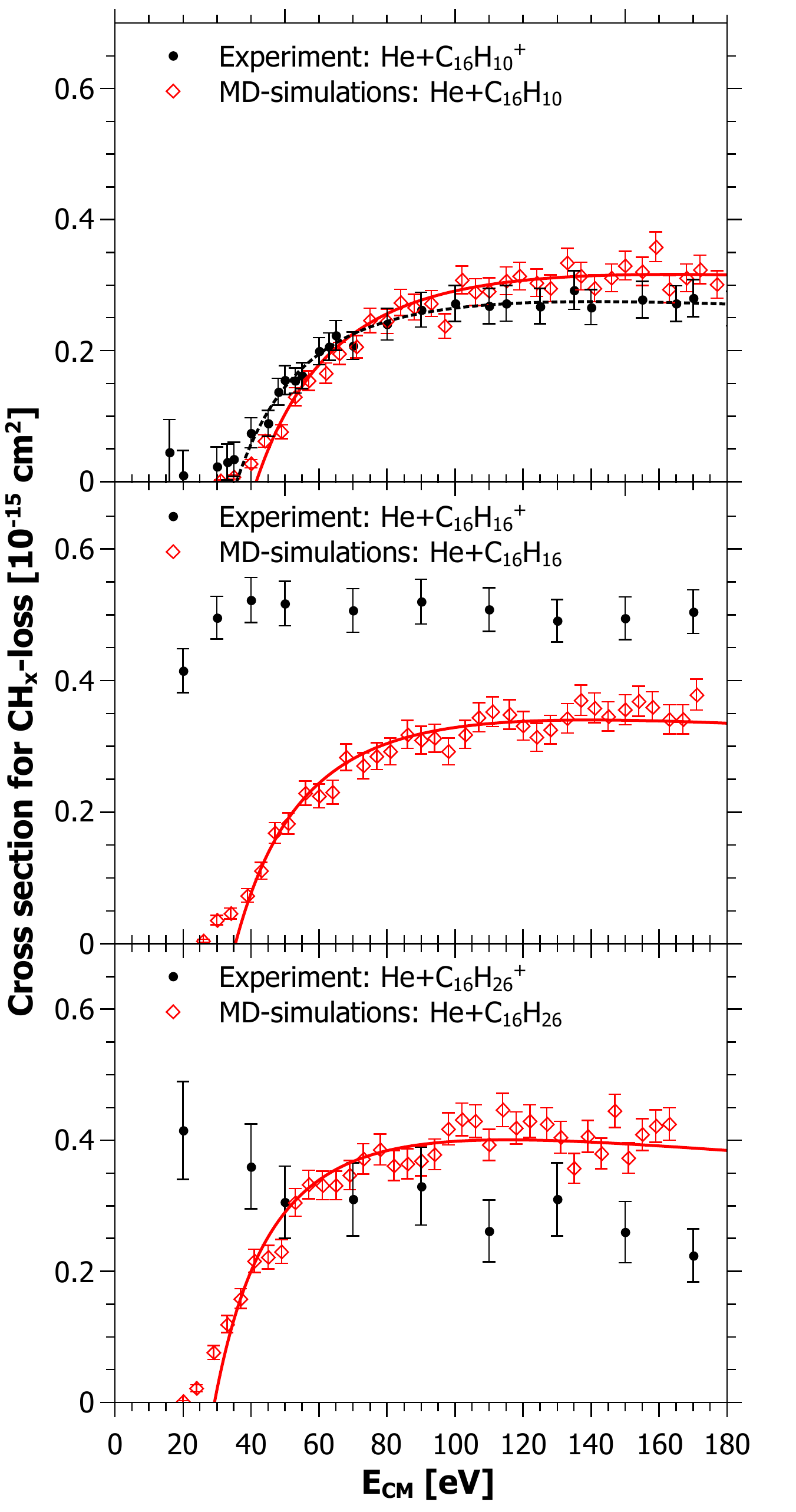}
}
\caption{
Experimental (filled symbols) absolute cross sections for CH$_x$-loss in He-C$_{16}$H$_{10+m}^+$ collisions as functions of center-of-mass energy $E_{CM}$. The corresponding MD-results are shown as open symbols for $m=0$ (top panel), $m=6$ (middle), and $m=16$ (bottom). Note that the simulated cross sections do not include effects of secondary fragmentation following the initial CH$_{x}$ knockout. For hydrogenated pyrene, the experimental cross sections leading to CH$_x$-loss are dominated by \textit{statistical} fragmentation processes (middle and lower panel), while \textit{non-statistical} (knockout) fragmentation dominates CH$_x$-loss for native pyrene (top panel). The solid and dashed lines are fits to Eq. (\ref{equ: fit}) for the MD and C$_{16}$H$_{10}^+$-experimental (top panel) data, respectively.}
\label{fig: cross_sections} 
\end{figure}

In Fig. \ref{fig: cross_sections} we show comparisons between cross sections from classical MD simulation for direct single carbon knockout for the neutral molecules and the present experimental CH$_x$-loss cross sections for the corresponding cations on an absolute scale. The experimental cross sections for each center-of-mass energy are deduced from the area under the n$_C = 15$-peaks in the mass spectra. The curves in Fig. \ref{fig: cross_sections} are fits to the analytical expression for the carbon knockout cross section
\begin{equation}
\sigma _{KO} = \frac{A/E_{CM}}{\pi^2 \arccos^{-2}(\sqrt{E_{th}/E_{CM}})-4}
\label{equ: fit}
\end{equation} 
as given in Ref. \cite{Stockett2015}. The factor $A$ and the threshold energy for knockout, $E_{th}$, are treated as free parameters. The threshold energies for the best fits are shown in Table \ref{tab:1} for the MD simulations (all three molecules in Fig. \ref{fig: pyrene}) and for the experimental results (C$_{16}$H$_{10}^{+}$ only). 
The displacement energy for an isolated PAH molecule is defined as the energy transfer at the knockout threshold. In our case, when we study collisions between He atoms and PAH cations this can also be taken as the energy loss of the He atom, $\Delta E_\text{He}$, at this particular energy. In Table \ref{tab:1}, we give Semi-Empirical (SE) displacement energies, $T^{SE}_{disp}$, which are based on fits of Eq. (\ref{equ: fit}) to the present experimental data and MD-simulations of $\Delta E_\text{He}$ as functions of He-PAH center-of-mass collision energy (cf. Fig. 4 in Ref. \cite{Stockett2015}). We also give displacement energies that are fully based on the MD-simulations, $T^{MD}_{disp}$, in Table \ref{tab:1}. There, we have fitted the expression in Eq. (\ref{equ: fit}) to the absolute MD-cross section for CH$_x$-loss and used it together with the MD result for the energy loss of the He atom. 

The threshold and displacement energies from the MD simulations decrease with increasing degree of hydrogenation. This is due to the weakening of the C-C bonds. The experimental results for C$_{16}$H$_{10}^+$ \cite{Stockett2015} (top panel, Fig \ref{fig: cross_sections}) compare favourably with the MD simulations for single carbon knockout from C$_{16}$H$_{10}$. This shows that CH$_x$-loss is indeed mainly the result of prompt knockout and that the cross section for this process is negligible below some threshold energy \cite{Stockett2015}. Statistical fragmentation does not seem to play a role for CH$_x$-loss from C$_{16}$H$_{10}$, which is consistent with calculations of a higher dissociation energy for CH-loss (7.10~eV) than for H- (5.16~eV) and C$_2$H$_2$-loss (6.30~eV) from Ref. \cite{Gatchell2015}. In addition, single-carbon loss seems to be associated with high fragmentation barriers as suggested by the large displacement energies in Table \ref{tab:1}.  

In the middle and lower panels of Fig. \ref{fig: cross_sections} we show the results for C$_{16}$H$_{16}^{+}$/C$_{16}$H$_{16}$ and C$_{16}$H$_{26}^{+}$/C$_{16}$H$_{26}$, respectively. Here, the lowest dissociation energies correspond to those for CH$_3$-loss at 2.26~eV and 1.60~eV for C$_{16}$H$_{16}^{+}$ and C$_{16}$H$_{26}^{+}$, respectively \cite{Gatchell2015}. The fact that our measured cross sections for CH$_x$-loss do not decrease below the MD-thresholds for knockout for C$_{16}$H$_{16}^+$ and C$_{16}$H$_{26}^+$ strongly indicates that these fragmentation processes mainly are statistical. 
When comparing the absolute MD-simulation and experimental CH$_x$-loss cross sections, one should note that the former only includes non-statistical fragmentation while the latter can also include statistical fragmentation. This results in the discrepancies visible in the absolute values for the hydrogenated molecules.


\subsection{Internal energy distributions}

\begin{table}
\caption{Single carbon knockout thresholds obtained through fits of Eq. (\ref{equ: fit}) to the experimental and MD simulation results in Fig. \ref{fig: cross_sections}. The experimental results for pyrene (C$_{16}$H$_{10}^+$) are based on data from Ref. \cite{Stockett2015}. Shown are also Semi Empirical (SE) and MD displacement energies---the energy losses of the He atom at the experimental and MD knockout thresholds \cite{Gatchell2015}.}
\label{tab:1}       
\begin{tabular}{l|c|c|c|c}
PAH & $E_{\text{th}}^{\text{Exp}}$ (eV)& $E_{\text{th}}^{\text{MD}}$ (eV)& $T_{\text{disp}}^{\text{SE}}$ (eV)& $T_{\text{disp}}^{\text{MD}}$ (eV)     \\
\noalign{\smallskip}\hline\noalign{\smallskip}
C$_{16}$H$_{10}$ & $35.8 \pm 0.9$  & $41.6 \pm 0.6 $ & $26.4 \pm 0.5 $ & $28.8 \pm 0.3$ \\
\noalign{\smallskip}\hline\noalign{\smallskip}
C$_{16}$H$_{16}$  & - & $35.4 \pm 0.7$ & - & $24.9 \pm 0.4$ \\
\noalign{\smallskip}\hline\noalign{\smallskip}
C$_{16}$H$_{26}$  & - & $29.3 \pm 0.7$ & - & $21.1 \pm 0.4$ \\
\noalign{\smallskip}\hline
\end{tabular}
\end{table}

\begin{figure*}
\centering
\resizebox{0.85\textwidth}{!}{%
\includegraphics{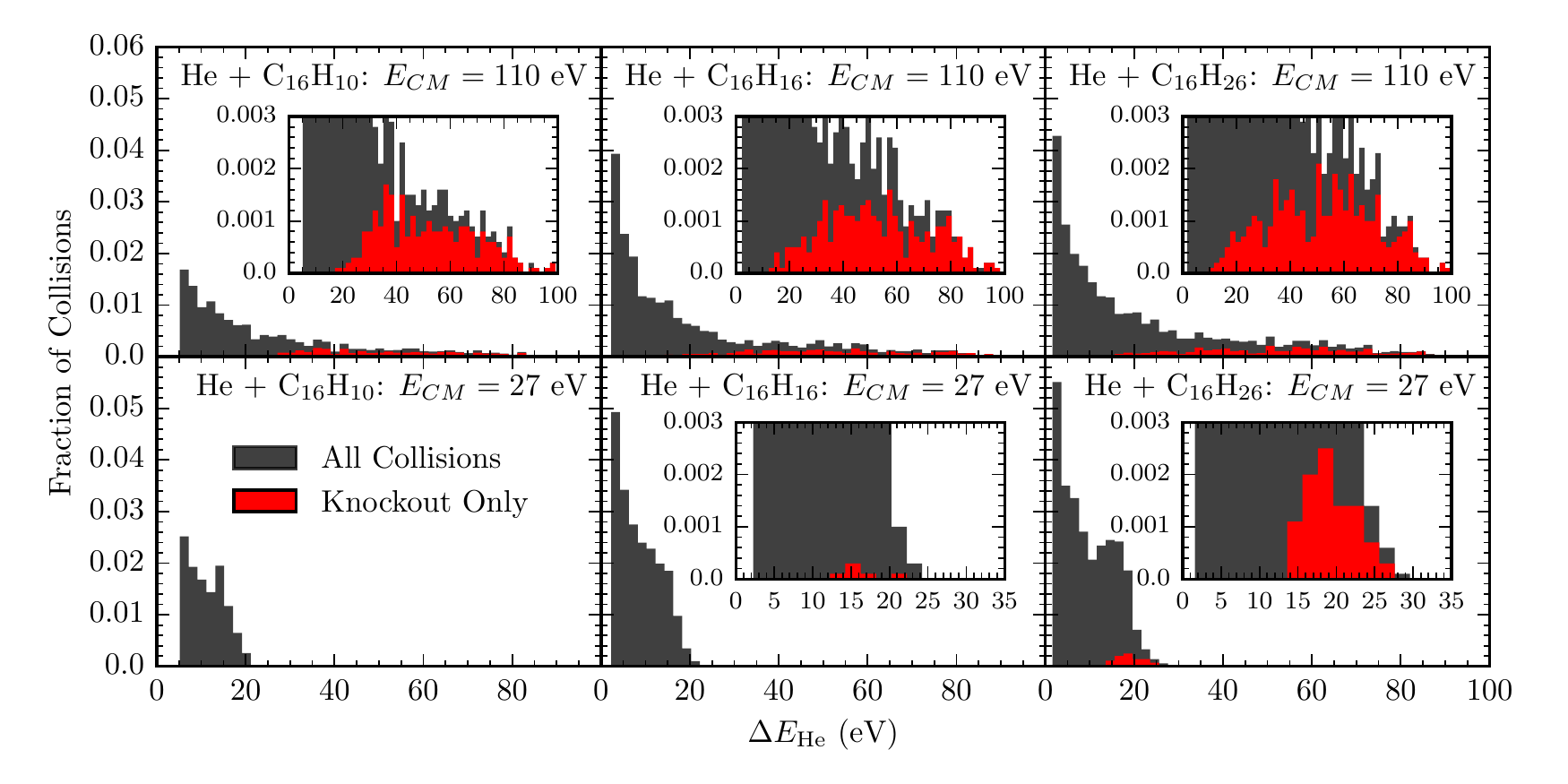}
}
\caption{Distributions of the energy transfers to the molecular systems in collisions between He and C$_{16}$H$_{10+m}$ ($m$ = 0, 6, 16) at $E_{CM} = 110$~eV (top row) and $E_{CM} = 27$~eV (bottom row) from our MD simulations. Insets show zoom-ins of the subset of collisions that lead to the prompt knockout of one C atom (in red). The distributions only show results for collisions with energy transfers above the lowest dissociation channel of the cation. These channels are H-loss (5.16~eV, C$_{16}$H$_{10}^{+}$), CH$_3$-loss (2.26~eV, C$_{16}$H$_{16}^{+}$) and CH$_3$-loss (1.60~eV, C$_{16}$H$_{26}^{+}$) \cite{Gatchell2015}.}
\label{fig: energy_dist}       
\end{figure*}

In Fig. \ref{fig: energy_dist} we show our MD results for the energy transfers to native (left panels) and hydrogenated (middle and right panels) pyrene molecules following collisions with He at center-of-mass energies of 110~eV (top row) and 27~eV (bottom row). The black bars show the full energy-transfer distributions, while the red bars show the parts of these distributions that lead to prompt single carbon knockout. Included are only results for collisions where energy equal to or above the lowest dissociation channel of the cation is transferred.  
Prompt carbon knockout from C$_{16}$H$_{10}$ occurs at 110~eV collision energy but not at 27~eV as can be seen in the leftmost panels of Fig. \ref{fig: energy_dist}. In both of these cases, however, energies well above the roughly 5~eV required for statistical fragmentation (H- and C$_2$H$_2$-loss) are often transferred. 
For C$_{16}$H$_{16}^{+}$, the lowest energy dissociation channel is CH$_3$-loss at 2.26~eV \cite{Gatchell2015}. This gives a much higher probability for statistical fragmentation than in the C$_{16}$H$_{10}$/ C$_{16}$H$_{10}^+$ case. That is, a much larger fraction of all collisions leads to internal energies above the lowest dissociation energy for C$_{16}$H$_{16}$/C$_{16}$H$_{16}^+$ than for C$_{16}$H$_{10}$/C$_{16}$H$_{10}^+$. This means that statistical fragmentation indeed should become much more likely for hydrogenated than for native pyrene. 

The two panels to the right in Fig. \ref{fig: energy_dist} show the distributions for C$_{16}$H$_{26}$, with the even lower lowest dissociation energy of the cation, 1.60~eV for CH$_3$-loss \cite{Gatchell2015}. The contribution from carbon knockout from collisions at 110~eV is again similar to the other molecules, but the fraction of events leading to single carbon knockout is significantly larger than for C$_{16}$H$_{10}$ and C$_{16}$H$_{16}$ at 27~eV. The fraction of collisions which transfer energies above the dissociation limit is even higher for C$_{16}$H$_{26}$ than for C$_{16}$H$_{16}$. The results for the hydrogenated molecules show that the energies transferred in the collisions with He are very similar to those for native pyrene---the main difference is that the dissociation energies are much lower for the hydrogenated molecules. This favours statistical fragmentation in these cases and makes CH$_x$-loss possible for collision energies far below the thresholds for knockout. 

\section{Summary and conclusions}

We have shown, by comparing experimental results with classical molecular dynamics simulations, that collision induced carbon backbone fragmentation is governed by very different processes when pyrene is hydrogenated and when it is not. For native pyrene individual carbon atoms may be knocked out promptly in fast non-statistical fragmentation processes. This may also occur for hydrogenated pyrene, but here it is also possible to break the carbon backbone through CH$_x$-loss channels with very low dissociation energies. This is shown explicitly through direct observations of significant CH$_x$-loss cross sections far below the carbon knockout thresholds at E$_{th}^{MD}=35.4 \pm 0.7$~eV and $29.3 \pm 0.7$~eV for C$_{16}$H$_{16}$ and C$_{16}$H$_{26}$, respectively. These are statistical fragmentation channels which should be dominant regardless of how the molecules are excited. Thus one can expect that hydrogenation would weaken PAHs also in photo-absorption processes. At least this would most likely be the case for pyrene. Remaining questions are if the situation is the same also for larger PAHs, like coronene, and how different numbers of added hydrogen atoms influence the stability.

Although the MD results suggest that prompt atom knockouts from hydrogenated PAHs are equally important first step processes as for fully aromatic PAHs we find no clear experimental fingerprints for this. The reason is that the so formed fragments do not survive on our experimental timescale of microseconds. Further it will also be difficult to directly observe non-statistical fragmentation in more complex systems where the bonds may have more of sp$^3$ characters such as e.g. biomolecules. However, it may be possible to observe such effects when biomolecules or hydrogenated PAHs are embedded in a surrounding (cluster) environment which may act as a cooling agent and a provider of building blocks for molecular growth processes \cite{Delaunay2015}. This calls for further studies of knockout driven fragmentation processes and secondary reactions.

\section{Acknowledgements}

This work was supported by the Swedish Research Council (Contracts Nos. 621-2012-3662, 621-2012-3660, 621-2014-4501, and 2015-04990). We acknowledge the COST action CM1204 XUV/X-ray Light and Fast Ions for Ultrafast Chemistry (XLIC).

\section{Author Contributions}

M.~Wolf and M.~Gatchell wrote the manuscript. The measurements were performed by M.~Wolf, L.~Giacomozzi, N.~de~Ruette, and M.~H.~Stockett. All authors were involved in the analysis of the results. The simulations were performed by M.~Gatchell and H.~Zettergren. H.~Cederquist, H.~T.~Schmidt and H.~Zettergren initiated the project. All authors have reviewed the manuscript.

%
 \bibliographystyle{epj}
 \bibliography{HPAH_Thresholds}
%
%
%

\end{document}